\documentclass[twocolumn]{aastex62}
\usepackage{amsmath,amstext}
\usepackage{comment}
\usepackage{apjfonts} 
\usepackage{chngcntr}

\newcommand{\be}{\begin{eqnarray}}
\newcommand{\ee}{\end{eqnarray}}
\newcommand{\lp}{\left(}
\newcommand{\rp}{\right)}
\newcommand{\lb}{\left[}
\newcommand{\rb}{\right]}

\begin{document}

\normalsize


\title{\Large \textbf{Shock Cooling Emission from Extended Material Revisited}}

\author{Anthony L. Piro}
\affiliation{The Observatories of the Carnegie Institution for Science, 813 Santa Barbara St., Pasadena, CA 91101, USA; piro@carnegiescience.edu}

\author{Annastasia Haynie}
\affiliation{Department of Physics and Astronomy, University of Southern California, Los Angeles, CA 90089, USA}
\affiliation{The Observatories of the Carnegie Institution for Science, 813 Santa Barbara St., Pasadena, CA 91101, USA; piro@carnegiescience.edu}

\author{Yuhan Yao}
\affiliation{Cahill Center for Astrophysics, California Institute of Technology, MC 249-17, 1200 E California Boulevard, Pasadena, CA 91125, USA}

\begin{abstract}
Following shock breakout, the emission from an astrophysical explosion is dominated by the radiation of shock heated material as it expands and cools, known as shock cooling emission (SCE). The luminosity of SCE is proportional to the initial radius of the emitting material, which makes its measurement useful for investigating the progenitors of these explosions. Recent observations have shown some transient events have especially prominent SCE, indicating a large radius that is potentially due to low mass extended material. Motivated by this, we present an updated analytic model for SCE that can be utilized to fit these observations and learn more about the origin of these events. This model is compared with numerical simulations to assess its validity and limitations. We also discuss SNe~2016gkg and 2019dge, two transients with large early luminosity peaks that have previously been attributed to SCE of extended material. We show that their early power-law evolution and photometry are well matched by our model, strengthening support for this interpretation.
\end{abstract}


\section{Introduction}
\label{sec:introduction}

Observations of supernovae (SNe) and other transients during the first few days provide valuable information about their progenitors \citep{Piro13}. Current and forthcoming surveys are making this an ideal time to study these early properties. The first electromagnetic emission is the shock breakout~(SBO) when the shock reaches an optical depth of $\sim c/v$, where $v$ is the shock speed \citep[][and references therein]{Waxman17}. SBO is short lived and high energy, and thus has only been seen in a handful of cases \citep[e.g.,][]{Soderberg08,Gezari15}. {In the hours to days following SBO, the hot shock-heated material expands and cools, giving rise to shock cooling emission \citep[SCE,][]{Grasberg76,Falk77,Chevalier92,Nakar10,Piro10,Rabinak11}. The timescale and temperature of SCE makes it well-suited for ground based optical observatories.}

As SCE has been detected more regularly, it has often been found to be more prominent than expected from typical blue or red supergiant models. This was first noticed for a subclass of SNe~IIb that show double-peaked light curves, such as SNe~1993J, 2011dh, 2011fu, and 2013df \citep{Wheeler93,Arcavi11,Kumar13,VanDyk14}. It is now generally accepted that their first peaks are due to SCE from low mass ($\sim0.01-0.1\,M_\odot$) extended ($\sim10^{13}\,{\rm cm}$) material \citep{Woosley94,Bersten12,Nakar14}. This unique structure is also reflected in pre-explosion imaging, which identified the progenitors as yellow supergiants \citep{Aldering94,Maund11,VanDyk11,VanDyk14}.

More recently, similar structures involving extended material have been invoked to explain a wide variety of transient events \citep[e.g.,][]{De18,Taddia18,Fremling19,Ho20,Jacobson20,Yao20}. Constraining the mass and radius of extended material usually involves comparing the observations to rough analytic scalings or fitting semi-analytic models \citep{Nakar14,Piro15,Nagy16,Sapir17}. The typical approach taken in developing these models is to begin with a density profile related to the physical conditions of the extended material (e.g., convective, radiative, wind-like) and introduce a shock velocity profile \citep{Matzner99} to understand the shock energy and ejecta velocity as a function of depth.

Here we explore a different approach in which we focus on the velocity profile of extended material once it is in the homologous phase. At such times, the ejecta naturally has a two component profile \citep{Chevalier89}, consisting of outer material with a strong velocity gradient and inner material with a more modest velocity gradient.
The advantage of this approach over \citet{Piro15} is that it better matches the expected properties of early SCE when the luminosity is  generated by the outermost material. We discuss useful rules of thumb for assessing whether SCE from extended material is appropriate for explaining a particular observation.

In Section~\ref{sec:framework}, we present an outline of our analytic model for SCE. We assess the validity of the model in Section~\ref{sec:numerical} by comparing to numerical calculations. In Section~\ref{sec:observations}, we apply our SCE model to two specific SNe with the purpose of demonstrating how the main features of SCE identified here are exemplified by these explosive events. We conclude in Section~\ref{sec:conclusion} with a summary of our results and a discussion of future work. {In Appendix~\ref{sec:appendixa}, we discuss the outer density profile of explosion models in more detail, while in Appendix~\ref{sec:appendix}, we provide further comparisons between our analytic SCE model and numerical calculations.}

\section{General Framework}
\label{sec:framework}

We consider extended material with mass $M_e$ and radius $R_e$, which is imparted with an energy $E_e$ as the shock passes through it. The material then expands homologously with radius $r=vt$, where $v$ is different for each shell of material. In essence, one can think of this as a coordinate system where each layer in the exploding ejecta is labeled by its velocity. We use a density structure that is inspired by the work of \citet{Chevalier89}, in which the extended material is divided into an outer region with a steep radial dependence,
\be
    \rho_{\rm out}(r,t) = \frac{KM_e}{v_t^3t^3}\lp\frac{r}{v_tt}\rp^{-n},
    \label{eq:rho_out}
\ee
and an inner density with a shallower radial dependence
\be
    \rho_{\rm in}(r,t) = \frac{KM_e}{v_t^3t^3}\lp\frac{r}{v_tt}\rp^{-\delta}.
    \label{eq:rho_in}
\ee
Typical values are $n\approx10$ and $\delta\approx1.1$ \citep{Chevalier89}, {although we discuss possible values for $n$ in more detail in Appendix~\ref{sec:appendixa}.} We show below that an attractive feature of our SCE solutions is that the exact values of $n$ and $\delta$ do not drastically alter our results as long as $n\gg1$ and $\delta\gtrsim1$. The parameter $K$ is set by mass conservation,
\be
    K = \frac{(n-3)(3-\delta)}{4\pi (n-\delta)}.
    \label{eq:K}
\ee
For typical values of $n$ and $\delta$, $K=0.119$. The parameter $v_t$ is the transition velocity between the outer and inner regions. Using energy conservation, this is found to be
\be
    v_t = \lb \frac{(n-5)(5-\delta)}{(n-3)(3-\delta)}\rb^{1/2} \lp\frac{2E_e}{M_e}\rp^{1/2},.
    \label{eq:vt}
\ee
Although in detail there should be some minimum velocity at the base of the extended material, for this work we assume it is negligible in comparison to $v_t$.

The optical depth as a function of radius and time is
\be
    \tau(r,t) = \int_r^\infty \kappa \rho(r,t)dr,
\ee
where for this work we use a constant electron scattering opacity due to the hot temperatures during SCE. See \citet{Rabinak11} for a treatment that considers more complicated opacity scalings. The result of this integral for the outer material is
\be
    \tau(r,t)=\frac{\kappa KM_e}{(n-1)v_t^2t^2}
    \lp \frac{r}{v_tt} \rp^{-n+1}.
    \label{eq:tau}
\ee
Setting the photosphere to be the depth where $\tau=2/3$, the photospheric radius evolves as
\be
    r_{\rm ph}(t) = (t_{\rm ph}/t)^{2/(n-1)} v_tt, \quad t\leq t_{\rm ph},
    \label{eq:rph1}
\ee
where
\be
    t_{\rm ph} = \lb \frac{3\kappa KM_e}{2(n-1)v_t^2}\rb^{1/2},
    \label{eq:tph}
\ee
is the time when the photosphere reaches the depth where the velocity is $v_t$. After this time, the photospheric radius should show a break in its evolution and decline more steeply. To solve for this, we find the $\tau=2/3$ depth by integrating through both the outer and inner material,
\be
    r_{\rm ph}(t)
    = \lb
        \frac{\delta-1}{n-1}
        \lp\frac{t^2}{t_{\rm ph}^2}-1 \rp
        +1
    \rb^{-1/(\delta-1)}v_tt, \quad t\geq t_{\rm ph},
\ee
In the limit that $t\gg t_{\rm ph}$, this can estimated as
\be
    r_{\rm ph}(t) \approx
    \lp\frac{n-1}{\delta-1}\rp^{1/(\delta-1)}
     (t_{\rm ph}/t)^{2/(\delta-1)} v_tt,\quad t\gg t_{\rm ph},
    \label{eq:rph2}
\ee
for the photospheric evolution. Formally, this suggests that the photosphere moves quickly inward in radius at these times since $\delta\approx1.1$. The exact evolution depends sensitively on $\delta$ though, so it is difficult to predict $r_{\rm ph}(t)$ without fitting to simulations. Furthermore, the photospheric evolution can be more complicated because effects like recombination and radioactive heating begin to be important before $t_{\rm ph}$.

We estimate the initial thermal energy in each shell with velocity $v$ to be
\be
    E_{\rm th,0}(v) \approx \int_{vt}^\infty 4\pi r^2\rho (v/2)^2 dr
    =\frac{\pi KM_ev_t^2}{n-5}
    \lp \frac{v_t}{v}\rp^{n-5}.
\ee
The factor of $2$ scaling for the velocity takes into account that the initial velocity imparted by the shock is accelerated by pressure gradients before reaching homologous expansion \citep{Matzner99}. This material adiabatically cools as it expands with the scaling $E_{\rm th}\propto 1/r$. The outer material all roughly starts with a similar initial radius of $R_e$, so that
\be
    E_{\rm th}(v,t)\approx E_{\rm th,0}(v)\lp\frac{R_e}{vt}\rp.
\ee
An observer at time $t$ observes radiation from the diffusion depth where $3\tau\approx c/v$. Although $\tau\approx c/v$ is commonly used as an approximation \citep[e.g.,][]{Nakar10}, we include the factor of $3$ to better match numerical results \citep{Morozova16}. Using Equation~(\ref{eq:tau}), the velocity at this depth is
\be
    v_{d}(t)
    = (t_{d}/t)^{2/(n-2)} v_t, \quad t\leq t_{d},
\ee
where
\be
    t_{d} = \lb \frac{3\kappa KM_e}{(n-1)v_tc}\rb^{1/2},
    \label{eq:tdif}
\ee
is the time at which the diffusion reaches the depth where the velocity is $v_t$. The radius of the diffusion depth is
\be
    r_{d}(t) = v_{d}(t)t
    = (t_{d}/t)^{2/(n-2)} v_tt,
    \quad t\leq t_{d},
\ee
At any time $t$, an observer sees a luminosity
\be
    L(t) \approx \frac{E_{\rm th,0}(v_{d}(t))}{t}
    \lp\frac{R_e}{v_{d}(t)t}\rp.
\ee
Putting together the above expressions, this is simplified to
\be
    L(t) \approx \frac{\pi(n-1)}{3(n-5)}
    \frac{cR_ev_t^2}{\kappa}
    \lp\frac{t_{d}}{t}\rp^{4/(n-2)},\quad t\leq t_{d}.
    \label{eq:learly}
\ee
This is just the classic result that the SCE luminosity is proportional to the initial radius of the shock heated material \citep[e.g.,][]{Nakar10,Piro10,Piro13}, with the addition of a power-law dependence with time, $L\propto t^{-4/(n-2)}$, dictated by the outer velocity profile.

To understand how the luminosity evolves for times after $t_{d}$, one might try to solve for the velocity of the diffusion depth just as we did above for the outer material. This would result in $v_{d}(t)\propto t^{2/(2-\delta)}$, which has a positive exponent because $\delta\sim1.1$. Such a scaling would seemingly imply the diffusion depth has reversed direction and is now moving into shallower material. The problem is that this material cannot radiate much more because, by this time, it has already lost most of its thermal energy. Instead, what happens is that the diffusion depth stays roughly fixed at $r_{d}\approx v_tt$ where it continues to radiate and cool.

To understand how the luminosity from the layer at $v_t$ changes with time, we solve the differential equation for the thermal energy of a radiation dominated gas subject to radiative cooling and adiabatic expansion \citep[e.g.,][]{Piro15}
\be
    \frac{dE_{\rm th}(v_t,t)}{dt} = -L - \frac{E_{\rm th}(v_t,t)}{t},
    \label{eq:energy equation}
\ee
where the radiative luminosity is
\be
    L = tE_{\rm th}(v_t,t)/t_{d}^2.
    \label{eq:luminosity}
\ee
We integrate Equation~(\ref{eq:energy equation}) from $t_{d}$ to $t$ to solve for $E_{\rm th}(v_t,t)$ and then substitute this result into Equation~(\ref{eq:luminosity}) to find
\be
    L(t) = \frac{E_{\rm th}(v_t,t_{d})}{t_{d}}
    \exp \lb -\frac{1}{2}\lp\frac{t^2}{t_{d}^2}-1 \rp \rb,
    \quad t\geq t_{d},
    \label{eq:llate}
\ee
where the prefactor is
\be
    \frac{E_{\rm th}(v_t,t_{d})}{t_{d}}
    = \frac{\pi(n-1)}{3(n-5)} \frac{cR_ev_t^2}{\kappa},
\ee
which matches onto the luminosity at $t\leq t_{d}$ given by Equation~(\ref{eq:learly}). This result shows that because the luminosity is mostly originating from a single layer at $r_{d}\approx v_tt$, it falls exponentially (rather than a power law like for the outer material). This is similar to the conclusion of \citet{Piro15}, who presents a one-zone treatment of SCE.

Following the exponential drop, heating from interior regions of the explosion (e.g., radioactive powering) will begin competing with the SCE. This will typically occur before the time $t_{\rm ph}$ because by the time the photosphere has passed through the depth at $v_t$, the extended material will have lost most of its thermal energy.

SCE roughly radiates as a black body (although see \citealp{Sapir17} for a more detailed discussion of thermalization) and thus the observed temperature is estimated as
\be
    T_{\rm BB} =
    \lp\frac{L}{4\pi r_{\rm ph}^2\sigma_{\rm SB}}\rp^{1/4},
\ee
where $\sigma_{\rm SB}$ is the Stefan-Boltzmann constant. Just as for the luminosity, the evolution of $T_{\rm BB}$ will probably become more complicated before $t_{\rm ph}$ because of additional heating sources. Additionally, both the photospheric depth and observed temperature will be impacted by other physics, such as recombination, not included here. For these reasons, the early scaling for $T_{\rm BB}$ and the break in $r_{ph}$ at $t_{d}$ are the most robust characteristics to compare with observations.

\begin{figure}
\includegraphics[width=0.43\textwidth,trim=13cm 0.2cm 0.5cm 0.0cm]{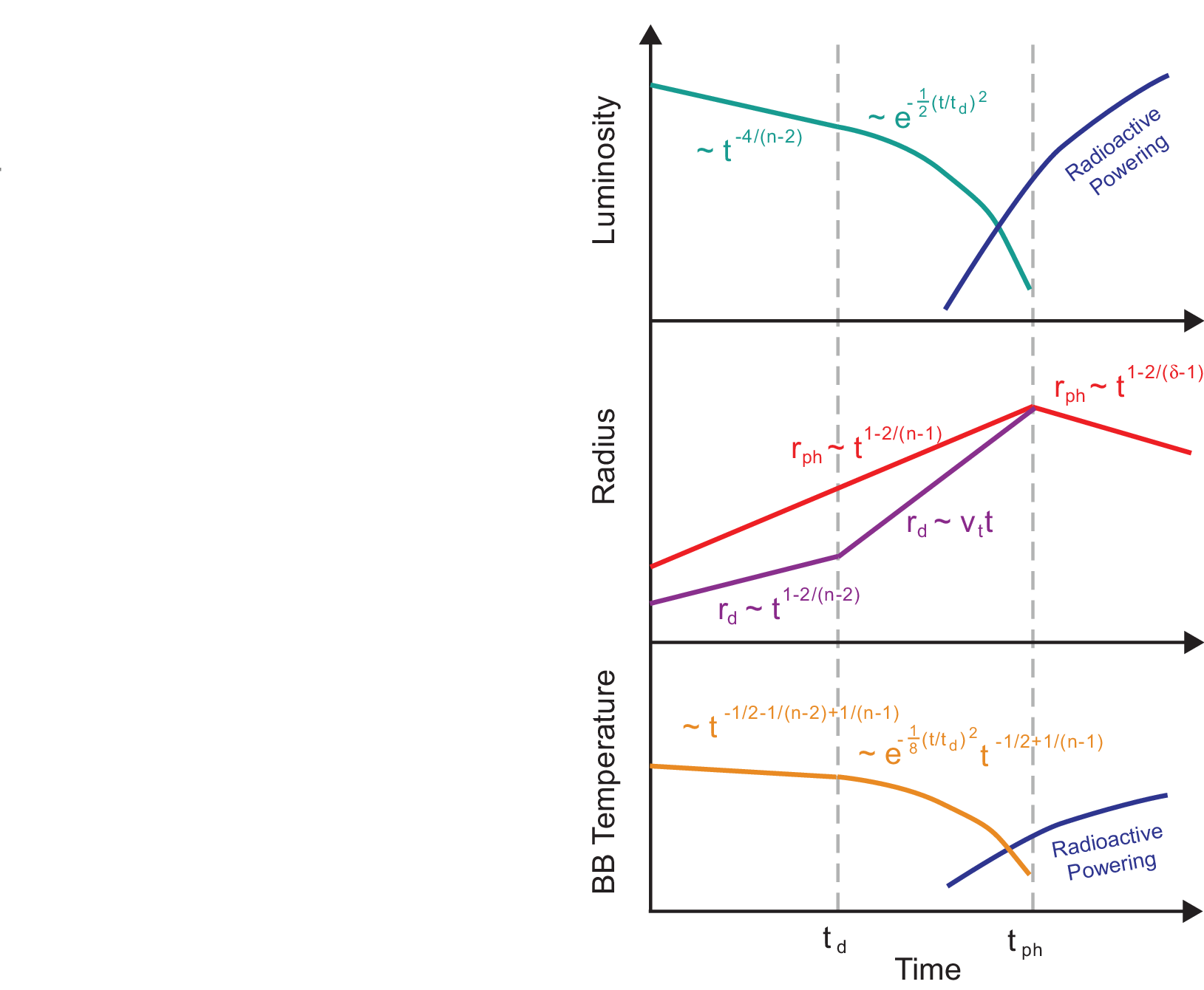}
\caption{Diagram showing the main phases of the luminosity (turquoise), photospheric radius (red), diffusion radius (purple), and black body temperature (orange) for SCE. The axis are logarithmic to emphasize the power-law time dependencies. The timescales of $t_{d}$ and $t_{\rm ph}$ correspond to where the luminosity and photospheric radius show breaks, respectively. At late times (but likely before $t_{\rm ph}$), the SCE luminosity and black body temperature evolution will be overtaken by radioactive heating (dark blue curves) and may not follow these scalings. These results also do not account for recombination, which can alter the depth of $r_{\rm ph}$ and $r_{d}$ and in turn also impact the luminosity and temperature.}
\label{fig:diagram}
\end{figure}

Figure \ref{fig:diagram} summarizes the main conclusions of the above discussion schematically. As SCE proceeds, the luminosity steepens from a power law to an exponential once the diffusion depth transitions from outer to inner material at time $t_{d}$. A key point is that as long as the opacity stays roughly constant, then the observed photospheric radius keeps the same power law even as the luminosity dropping faster. This evolution only changes at the later time of $t_{\rm ph}\approx (c/2v)^{1/2}t_d$, when the photosphere transitions into the inner material.

\section{Comparison to Numerical Models}
\label{sec:numerical}

We next turn to simulations to demonstrate how they roughly follow the properties described above. These calculations are similar to the Type IIb SN models of \citet{Piro17}, but we save discussing the full details until Appendix~\ref{sec:appendix}. The key points to mention here are that the models consist of a $3.55\,M_\odot$ helium core (once the inner $1.4\,M_\odot$ has been excised to represent formation of a neutron star) with hydrogen-rich extended material siting atop. For the particular example here we use $M_e=0.017\,M_\odot$ and $R_e=125\,R_\odot$.

These models are exploded with our open-source numerical code \texttt{SNEC} \citep{Morozova15} with an explosion energy of $E_{\rm SN}=10^{51}\,{\rm erg}$, but only a fraction of this energy makes its way into the extended material. This is estimated to be \citep{Nakar14}
\be
    E_e \approx 2\times10^{49}E_{51}
    \lp\frac{M_c}{3\,M_\odot}\rp^{-0.7}
    \lp\frac{M_e}{0.01\,M_\odot}\rp^{0.7}
    {\rm erg},
    \label{eq:e_e}
\ee
where $E_{51}=E_{\rm SN}/10^{51}\,{\rm erg}$ and $M_{c}$ is the mass of the helium core. For this case, we find \mbox{$E_e=2.6\times10^{49}\,{\rm erg}$.} Substituting this energy into Equation~(\ref{eq:vt}) results in $v_t=2.1\times10^9\,{\rm cm\,s^{-1}}$.

\begin{figure}
\includegraphics[width=0.45\textwidth,trim=0cm 6.5cm 1.5cm 0.0cm]{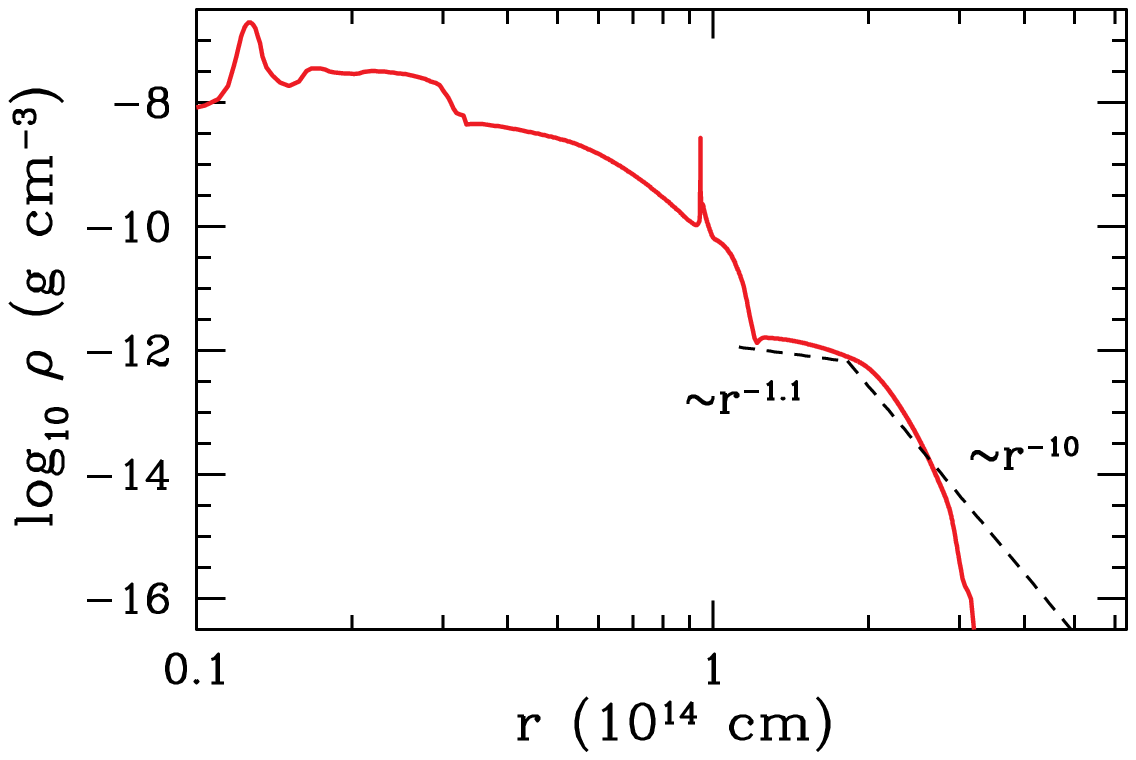}
\caption{Numerical density profile for a SN model taken $1\,{\rm day}$ after explosion (red curve). The strong break at a radius a little above \mbox{$10^{14}\,{\rm cm}$} roughly divides the interior helium core ($M_c=3.55\,M_\odot$) from the hydrogen-rich extended material ($M_e=0.017\,M_\odot$). The dashed line shows the analytic density profile using Equations~(\ref{eq:rho_out}) and (\ref{eq:rho_in}) with $v_t=2.1\times10^9\,{\rm cm}$.}
\label{fig:rho_radius}
\end{figure}

In Figure \ref{fig:rho_radius}, we present a snapshot of the density profile from this numerical model at $1\,{\rm day}$ following explosion (red curve). The strong density break above $\approx10^{14}\,{\rm cm}$ corresponds to the top of the helium core, above which the extended hydrogen-rich material sits. Although the overall density profile can be fairly complicated given the many different compositional layers within the star, the extended material clearly shows the shallow inner and steep outer density profile. The dashed line corresponds to the analytic density profile using \mbox{Equations (\ref{eq:rho_out}) and (\ref{eq:rho_in})} with $v_t=2.1\times10^9\,{\rm cm}$. {This comparison shows that the analytic density profile is a reasonable, albeit not exact, description of the numerical result. The largest discrepancy is at the outermost regions, but because the diffusion wave quickly passes through these regions, $\rho\propto r^{-10}$ ends up being a good approximation.}

\begin{figure}
\includegraphics[width=0.43\textwidth,trim=0.0cm 0cm 4.5cm 0.0cm]{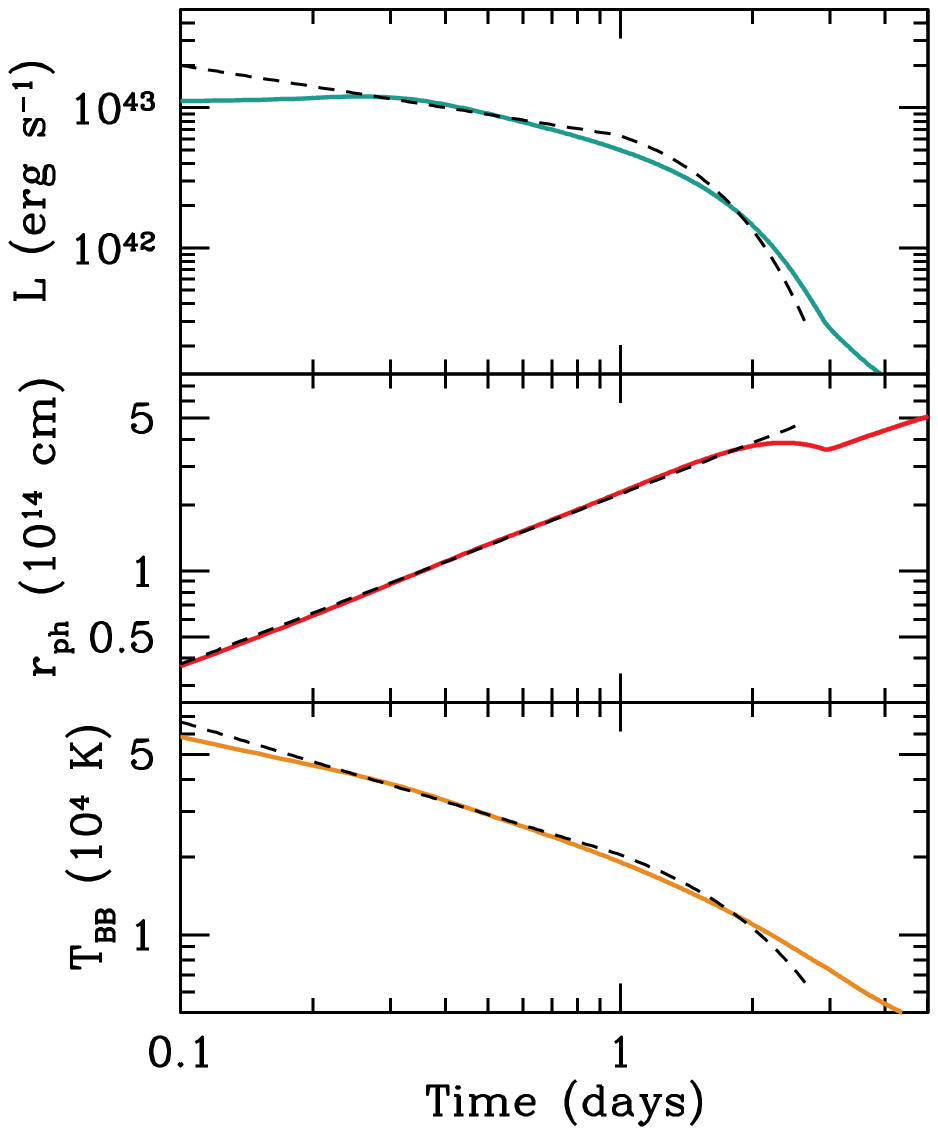}
\caption{Luminosity (turquoise), photospheric radius (red), and black body temperature (orange) from the numerical model used for the profile plotted in Figure \ref{fig:rho_radius}. The dashed lines show the analytic results for extended material with $M_e=0.017\,M_\odot$ and \mbox{$R_e=125\,R_\odot$} using $n=10$, $\delta=1.1$ and $v_t=2.1\times10^9\,{\rm cm\,s^{-1}}$.}
\label{fig:numerical}
\end{figure}

In Figure~\ref{fig:numerical}, we compare the observables of the simulations to our analytic model. In the top panel, we compare the SCE luminosity using Equations~(\ref{eq:learly}) and (\ref{eq:llate}) where $t_d=0.98\,{\rm days}$. The analytic solutions are plotted out to $t_{\rm ph}=2.63\,{\rm days}$, which roughly matches when the simulations begin diverting from the analytics due to radioactive heating. While the luminosity shows a break at $t_d$, the photospheric radius continues as a power law that closely follows the evolution predicted by Equation~(\ref{eq:rph1}) all the way up until $t_{\rm ph}$. At this point, the photosphere even decreases in radius as found in Equation~(\ref{eq:rph2}), but the time dependence of $r_{\rm ph}$ for $t\gg t_{\rm ph}$ is so strongly dependent on $\delta$ that the analytics do not have much predictive power at these times.

Nevertheless, the takeaway from this comparison is that even if the density profile is only roughly replicated by the analytic model, the observables are reproduced fairly robustly. This provides some confidence that as these models are compared with observations, at least the main physical parameters will be constrained with some fidelity.

\section{Comparison to Observations}
\label{sec:observations}

We next compare our analytic results to a few observations where the presence of extended material and associated SCE has been claimed. This demonstrates how the observations show the main features we identify here, strengthening the interpretation of SCE.

\subsection{SN 2019dge}

SN~2019dge was a helium-rich supernova with a fast-evolving light curve indicating a low ejecta mass \citep{Yao20}. Its early rise was too rapid to be explained as radioactively powered diffusion, and was thus interpreted as being due to SCE from extended material.

\begin{figure}
\includegraphics[width=0.45\textwidth,trim=0.5cm 0.3cm 1.5cm 0.0cm]{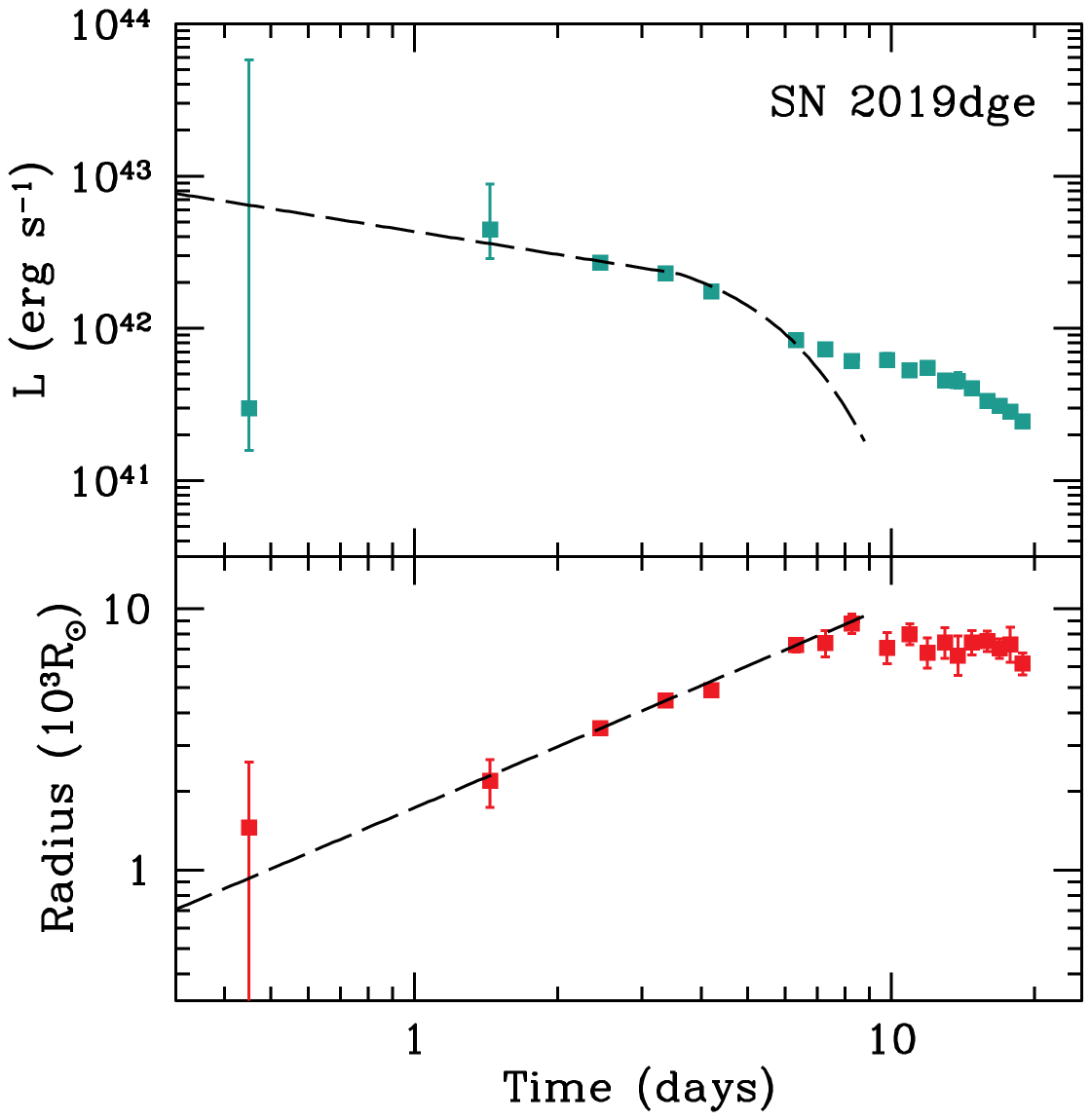}
\caption{Observed bolometric luminosity (turquoise) and photospheric evolution (red) of SN~2019dge \citep{Yao20}. Dashed lines are from our analytic model using $M_e=0.14\,M_\odot$, $R_e=205\,R_\odot$, and \mbox{$E_e=5.2\times10^{49}\,{\rm erg}$.}}
\label{fig:19dge}
\end{figure}

\begin{figure}
\includegraphics[width=0.45\textwidth,trim=0.3cm 0.3cm 1.7cm 0.0cm]{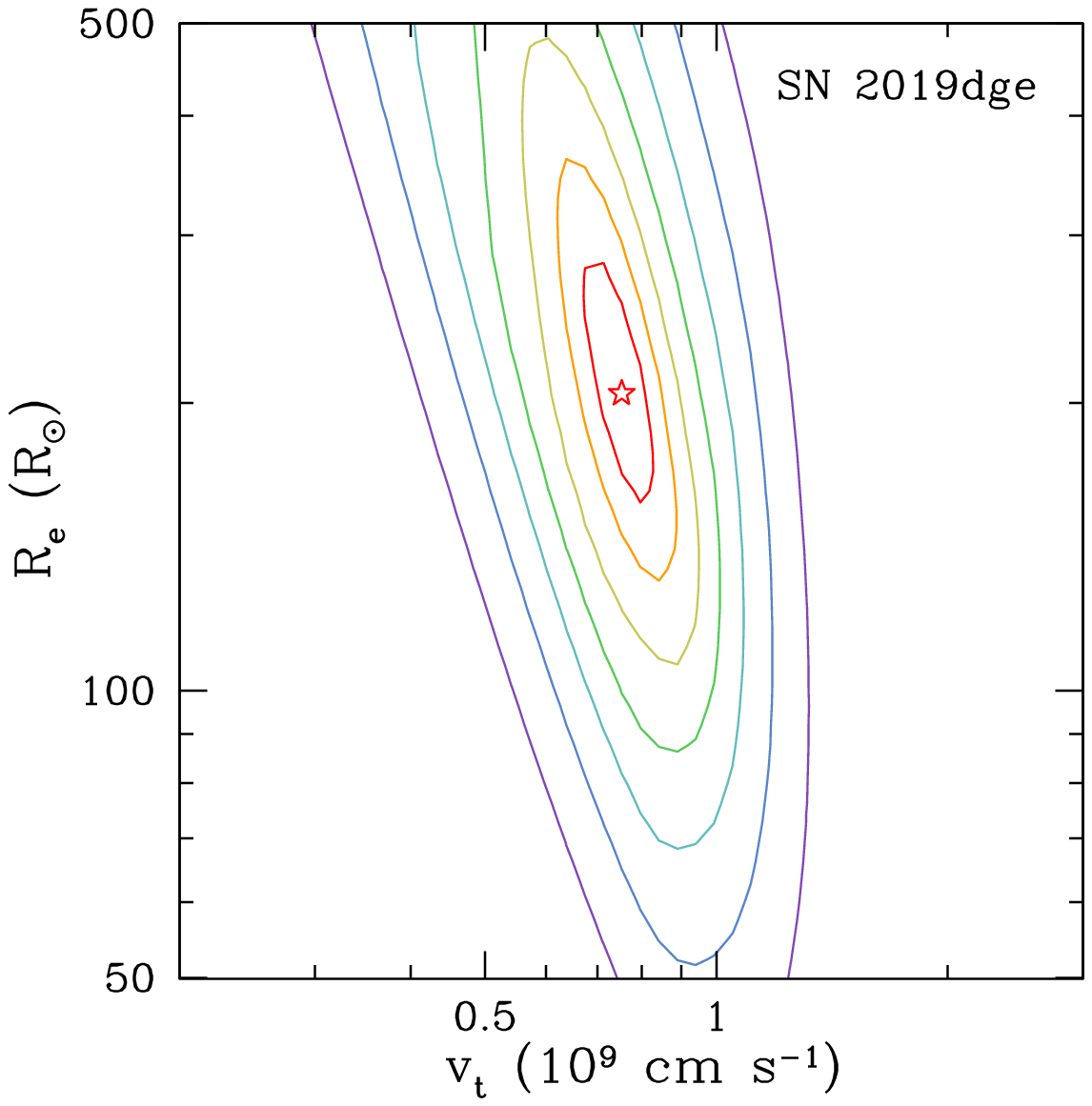}
\caption{Contours of constant $\log_{10}(\chi^2/\chi^2_{\rm min})$ as a function of $R_e$ and $v_t$ from fitting the first six data points of SN~2019dge. The red star marks the location of $\chi^2_{\rm min}$ with $M_e=0.14\,M_\odot$, $R_e=205\,R_\odot$, and $v_t=7.5\times10^8\,{\rm cm\,s^{-1}}$. The contours correspond to $\log_{10}(\chi^2/\chi^2_{\rm min})=0.25$ to $1.75$ spaced by intervals of $0.25$.}
\label{fig:chisq_19dge}
\end{figure}

An important difference between \citet{Piro15} and the updated model here is that we now include the power-law evolution of $L(t)$ for $t<t_d$ as well as a more careful treatment of the scaling of $r_{\rm ph}$. It is therefore interesting to investigate whether SN~2019dge shows such evolution since most other SNe do not have such detailed early observations. In Figure~\ref{fig:19dge}, we plot the observed bolometric luminosity and photospheric radius using logarithmically spaced coordinates. Presenting the data in this way immediately makes the power-law evolution of these quantities clear in a way that may not be as obvious if the coordinates were plotted linearly.

We fit the first six data points of $L$ and $r_{\rm ph}$ using our analytic model and evaluating $\chi^2$. Motivated by the helium-rich composition of this event, we set $\kappa=0.2\,{\rm cm^2\,g^{-1}}$. Contours of constant $\log_{10}(\chi^2/\chi^2_{\rm min})$ are plotted in Figure~\ref{fig:chisq_19dge} as a function of $R_e$ and $v_t$. The red star marks the location of $\chi^2_{\rm min}$. The best fitting value of $M_e$ does not change much, ranging from $M_e\approx0.1-0.3\,M_\odot$ over this parameter space. To better understand how these parameters are set by the observations, note that $t_d$ is well constrained by when the luminosity begins to drop, which puts constraints on the ratio $(M_e/v_t)^{1/2}$. Combining this with the normalization of $r_{\rm ph}(t)$, which scales as $M_e^{1/9}v_t^{7/9}$ for $n=10$, allows us to independently constrain $M_e$ and $v_t$. Finally, using the value of $v_t$ we find, the overall normalization of $L(t)$ constrains $R_e$.

The best fitting model is plotted in Figure~\ref{fig:19dge} with parameters $M_e=0.14\,M_\odot$, $R_e=205\,R_\odot$, and $E_e=5.2\times10^{49}\,{\rm erg}$ (or equivalently $v_t=7.5\times10^8\,{\rm cm\,s^{-1}}$). If we were instead to use the model from \citet{Piro15} to fit this event, the result would be a much smaller radius (by a factor of $\sim5$) and a larger explosion energy. This is mostly due to the differences in the scaling of $r_{\rm ph}$ between the two models. In this updated work, correctly including the velocity gradient results in larger photospheric velocities. In contrast, since the work of \citet{Piro15} would predict smaller photospheric velocities, this must be compensated by inferring a larger energy to match the observed colors of SN~2019dge, which in turn implies a smaller radius since there is a degeneracy between $E_e$ and $R_e$ in the luminosity.

Nevertheless, the main point of this comparison is not the specific parameters but how the data clearly scales as expected from our analytic work. This provides even stronger support for the SCE interpretation. Looking for power-law behavior at early times will be a useful way to identify whether SCE is being seen in other future events.

\subsection{SN 2016gkg}

\begin{figure}
\includegraphics[width=0.44\textwidth,trim=0.0cm 0.3cm 1.3cm 0.5cm]{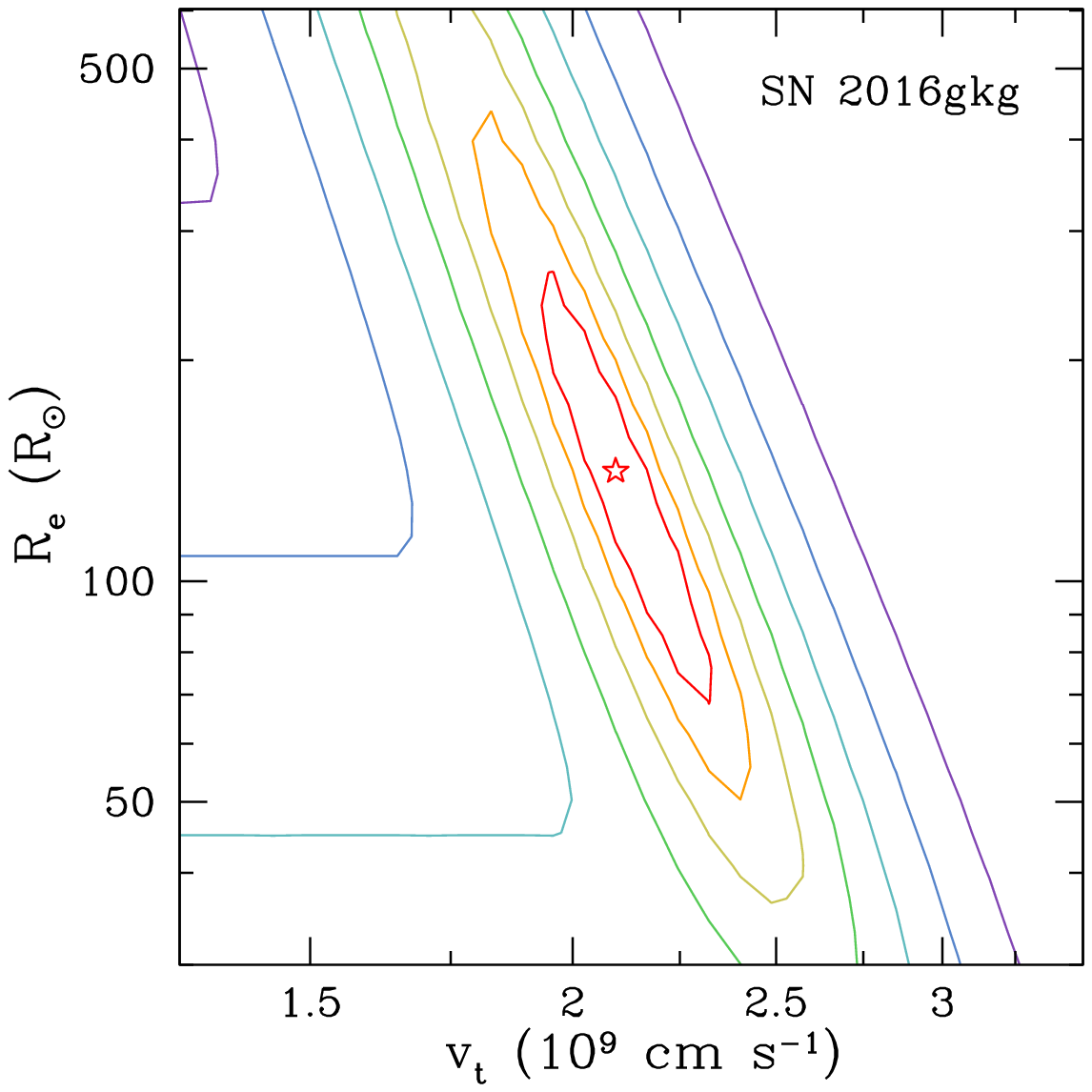}
\caption{Contours of constant $\log_{10}(\chi^2/\chi^2_{\rm min})$ as a function of $R_e$ and $v_t$ from fitting the SCE of SN~2016gkg. The red star marks the location of $\chi^2_{\rm min}$ with $M_e=0.03\,M_\odot$, $R_e=141\,R_\odot$, and $v_t=2.1\times10^9\,{\rm cm\,s^{-1}}$. Contours are spaced the same as Figure~\ref{fig:chisq_19dge}.}
\label{fig:chisq_16gkg}
\end{figure}

We next consider our model in comparison to SN~2016gkg. This was a Type IIb SN that was caught especially early after explosion and shows a prominent double-peaked light curve. It has well-sampled multi-band coverage including ultraviolet wavelengths \citep{Arcavi17,Kilpatrick17,Tartaglia17}. Especially unique is the amateur data found during the rise that greatly restricts the explosion time \citep{Bersten18}. Although early work constrained the radius of the extended material using a variety of analytic and semi-analytic work \citep[e.g.,][]{Rabinak11,Nakar14,Piro15,Sapir17}, a grid of numerical simulations was required to provide a detailed fit to the multi-band light curves of SCE \citep{Piro17}.

In Figure~\ref{fig:chisq_16gkg}, we plot contours of constant $\log_{10}(\chi^2/\chi^2_{\rm min})$ from fitting the first $\approx3.5\,{\rm days}$ of SN~2016gkg multi-band photometry. 
Note that the horizontal contours on the left side of the plot are due to issues with matching the early data points. The best fitting model is mostly consistent with \citet{Piro17}. The advantage of this approach is that we are able to consider a much wider range of models over a shorter period of time. More specifically, \citet{Piro17} considered $2,400$ models with varying $M_e$, $R_e$, and $E_e$, while we consider $250,000$ models in a fraction of the time. Our best fit model is slightly higher energy (and smaller radius) than \citet{Piro17}, and this was partially due to our ability to consider a wider range of possible parameters here.

\begin{figure}
\includegraphics[width=0.44\textwidth,trim=0.0cm 0.3cm 1.3cm 0.5cm]{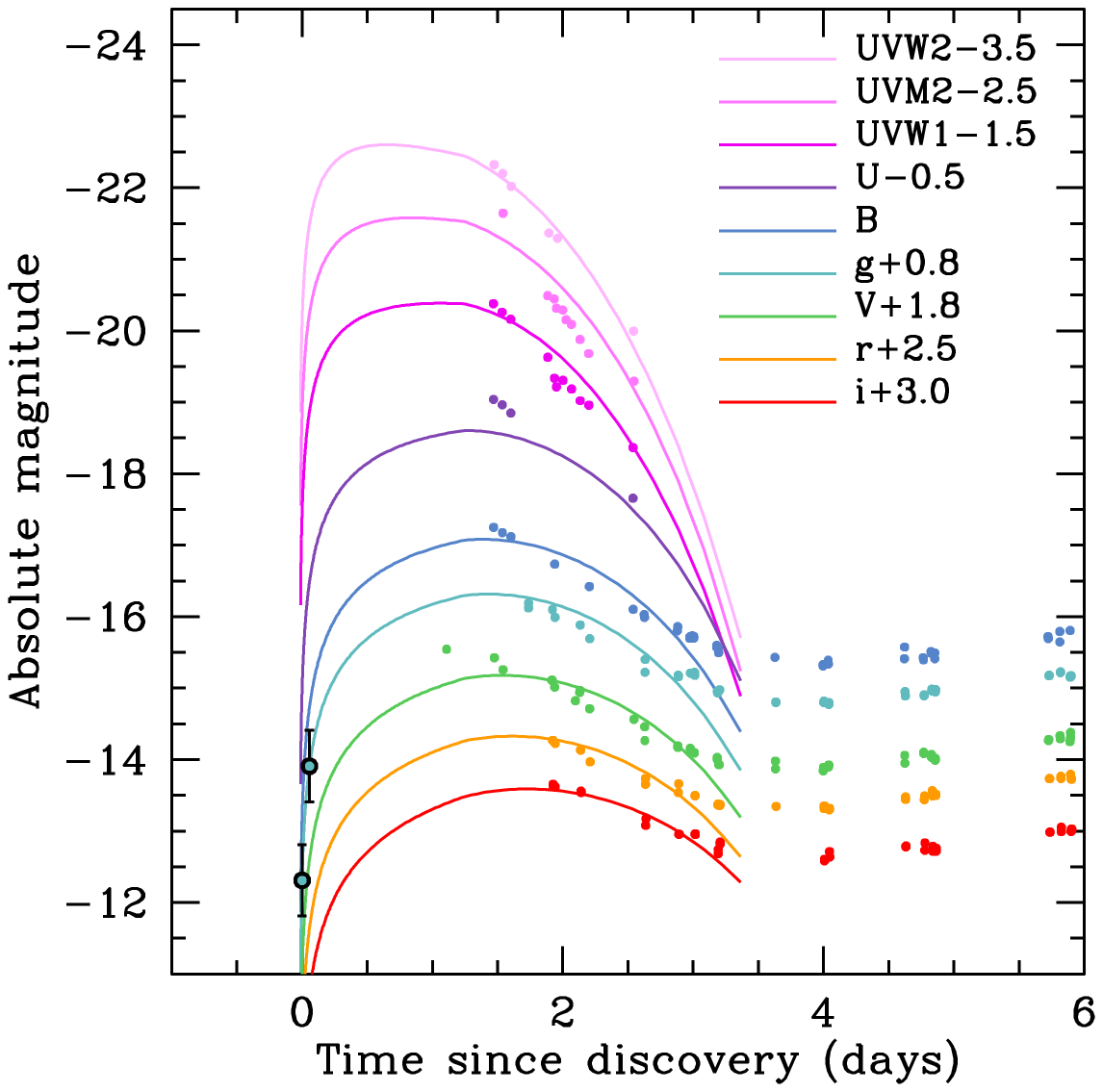}
\caption{Colored points are photometry from SN~2016gkg \citep{Piro17}, focusing on the first peak due to SCE. {Solid lines} are our analytic model using $M_e=0.03\,M_\odot$, $R_e=141\,R_\odot$ and \mbox{$E_e=8.3\times10^{49}\,{\rm erg}$.}}
\label{fig:16gkg}
\end{figure}

In Figure~\ref{fig:16gkg}, we plot the best fitting model in comparison to the multi-band photometry. The model presented here is especially good for fitting the early rise in comparison to the model by \citet{Piro15}, which has trouble fitting the earliest, bluest data as shown in \citet{Arcavi17}.

\section{Discussion and Conclusion}
\label{sec:conclusion}

In this work, we have presented an updated analytic model for SCE from extended material. Our approach is different from previous work in that we consider the ejecta once it has already reached the homologous phase to derive our results. This has the advantage of resolving the early, power-law evolution for the luminosity that is not addressed by \citet{Piro15} as well as improving our treatment of the photospheric evolution. This allows us to better match the earliest, bluest phases of SCE. We provide comparisons to numerical models and observations of SCE to demonstrate the advantage of this approach. The luminosity and photospheric radius evolution of SN~2019dge exhibit the scalings expected from our model, strengthening the support for a SCE interpretation.

The analytic model we present can be used to fit observations with the parameters $M_e$, $R_e$, and $v_t$ (or alternatively $E_e$). In principle, one can also fit for the outer density steepness $n$, but we show that the solutions are fairly insensitive to the exact value of $n$ as long as $n\sim10$. The main features of SCE of extended material are summarized as follows.
\begin{itemize}
\item The density is described as a steep outer profile and a shallow inner profile with a break at a transition velocity $v_t$.
\item The luminosity scales as $L\sim t^{-4/(n-2)}$ up to time $t_d$ when the diffusion wave reaches the depth where the velocity is $v_t$.
\item The diffusion depth is then roughly fixed at the depth of $v_t$. This causes the luminosity to drop exponentially from time $t_d$ up to a time $\sim t_{\rm ph}=(c/2v_t)^{1/2}t_d$.
\item The photospheric radius scales as $r_{\rm ph}\sim t^{1-2/(n-1)}$ and remains as roughly this power law even past time $t_d$.
\end{itemize}
Unfortunately, few observations have sufficient early data like SN~2019dge to resolve such scalings. For example, iPTF14gqr \citep{de18b} and iPTF16hgs \citep{De18} could in principle be ideal candidates for applying this theory, but they simply have insufficient early data to conclusively resolve if our predicted power laws are occurring. Our work provides strong motivation for high cadences at the earliest times to better test whether SCE from extended material is taking place. Multi-band coverage is also key so that a bolometric light curve can be reliably constructed.

In the future, there are a number of improvements that can be made to this this work. Here we fit the SCE component, but the second radioactively-powered peak should be fit simultaneously. Such a fit should also attempt to consistently resolve $E_e$ versus $E_{\rm SN}$. Here we relate these quantities using Equation (\ref{eq:e_e}), which is from \citet{Nakar14}, but this relation would benefit from calibration to numerical models.

Other details that can be improved are the relation between $n$ and the initial density profile of the extended material, as well as the treatment of the opacity. We consider it a strength of the model here that the results are fairly robust to uncertainties in $n$, nevertheless, as more information is known about the progenitor (for example, from pre-explosion imaging and numerical stellar model) it may be useful to consider a specific $n$ value other than $10$. For the opacity, we have focused on a constant value motivated by the high temperatures during SCE, but especially for helium-rich extended material, recombination may play a role which could make the outermost layers more transparent to electron scattering.

\acknowledgments
A.L.P. acknowledges financial support for this research from a Scialog award made by the Research Corporation for Science Advancement. A.H. acknowledges support from the USC-Carnegie fellowship. Y.Y. thanks the Heising-Simons Foundation for financial support.

\begin{appendix}
\counterwithin{figure}{section}

{\section{Analytic Exploration of the Shallow Density Gradient}\label{sec:appendixa}}

{In this work, we took the approach of considering the density profile well into the homologous phase, but another commonly used approach is to consider the initial velocity profile just following shock passage \citep[e.g.,][]{Piro10,Nakar10}. This work uses $v\propto \rho_0^{-\beta}$ for the shallowest layers \citep{Matzner99}, where $\rho_0$ is the density profile prior to explosion. Typically, $\beta\approx0.19$, which is fairly insensitive to the exact density profile. Here we explore what value of $n$ is implied for a certain value of $\beta$.}

{For the layers near the surface, their initial radii are all about the same at $\approx R_e$, and then following sufficient expansion, they reach a radius $r(\rho_0,t) \approx v(\rho_o)t$. Assuming an initial thickness of a layer $H_0(\rho_0)$, and using mass conservation, the density of the layer at any given time is
\be
    \rho(\rho_0,t) \approx \frac{R_e^2H_0(\rho_0)}{v(\rho_0)^3t^3}
    \rho_0.
\ee
The optical depth at any given time is
\be
    \tau(\rho_0,t)
    = \int \kappa \rho(\rho_0,t)dr
    \approx \frac{R_e^2H_0(\rho_0)}{v(\rho_0)^2t^2}
    \kappa \rho_0.
    \label{eq:taurho0}
\ee
From this we can solve for the photospheric velocity evolution, but this requires choosing $H_0(\rho_0)$. Generally speaking, for a polytropic index $s$, where typical values are $s=3$ for a radiative profile and $s=3/2$ for a convective profile (we use $s$ rather than the typical $n$ here since $n$ already has a different meaning), $H_0\propto \rho_0^{1/s}$ \citep{Piro10}. Substituting this into Equation (\ref{eq:taurho0}) and setting $\tau(\rho_0,t)=1$, we solve for the location of the photosphere with respect to the initial density of
\be
    \rho_{0,\rm ph}(t) \propto
    t^{2/(1+1/s+2\beta)},
\ee
so that
\be
    v_{\rm ph}(t) = v(\rho_{0,\rm ph}(t))
    \propto
    t^{-2\beta/(1+1/s+2\beta)}.
\ee
Comparing with the time dependence of $v_{\rm ph}=r_{\rm ph}/t \propto t^{-2/(n-1)}$ using Equation~(\ref{eq:rph1}), we find
\be
    n = (1+1/s+3\beta)/\beta.
\ee
Thus typical values are $n=10$ (for $s=3$) and $n=11.8$ (for $s=3/2$). Although these are roughly consistent with the value of $n$ we use in Section~\ref{sec:observations}, we also experimented by fitting SN 2016gkg using larger values of $n$. Our general result was that the fits gave roughly the same $M_e$, but $R_e$ was smaller by about $\sim10\%$. These models were also noticeably poorer at fitting the data, which suggests that $n=10$ may be more representative of what may be occurring in nature}

\section{Further Comparisons with Numerical Models}
\label{sec:appendix}

\begin{figure}
\includegraphics[width=0.335\textwidth,trim=0.0cm 6cm 0.0cm 0.0cm]{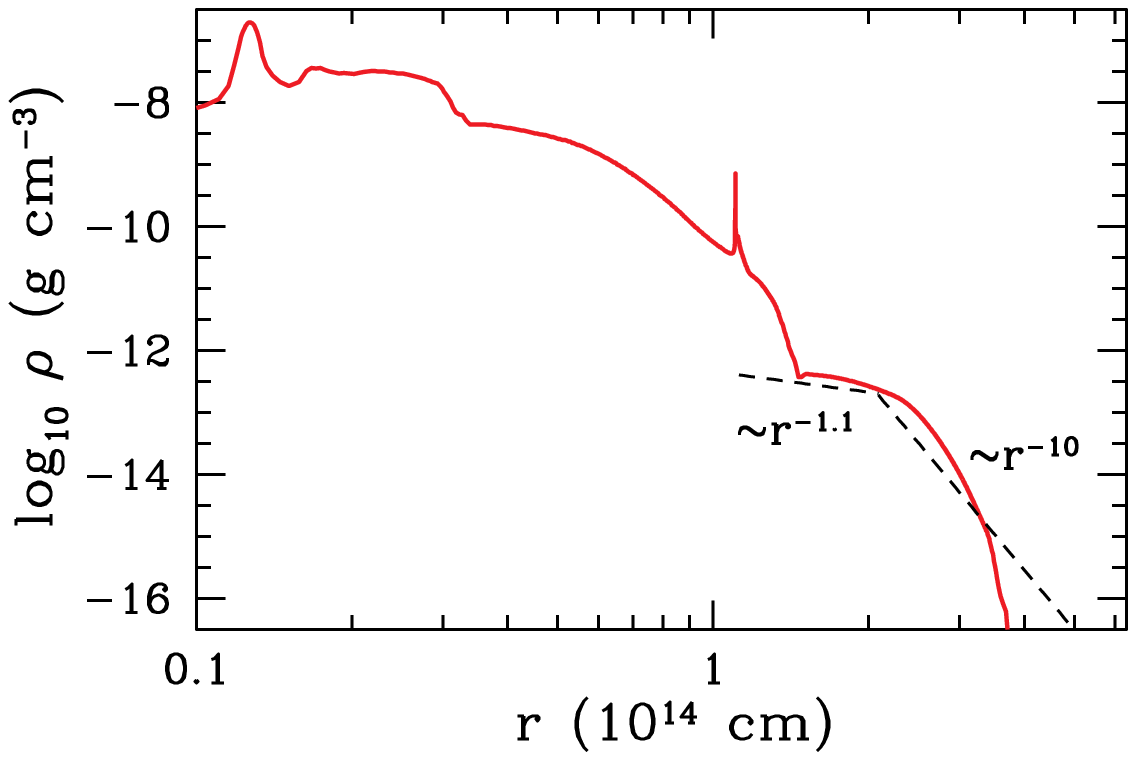}
\includegraphics[width=0.335\textwidth,trim=0.0cm 6cm 0.0cm 0.0cm]{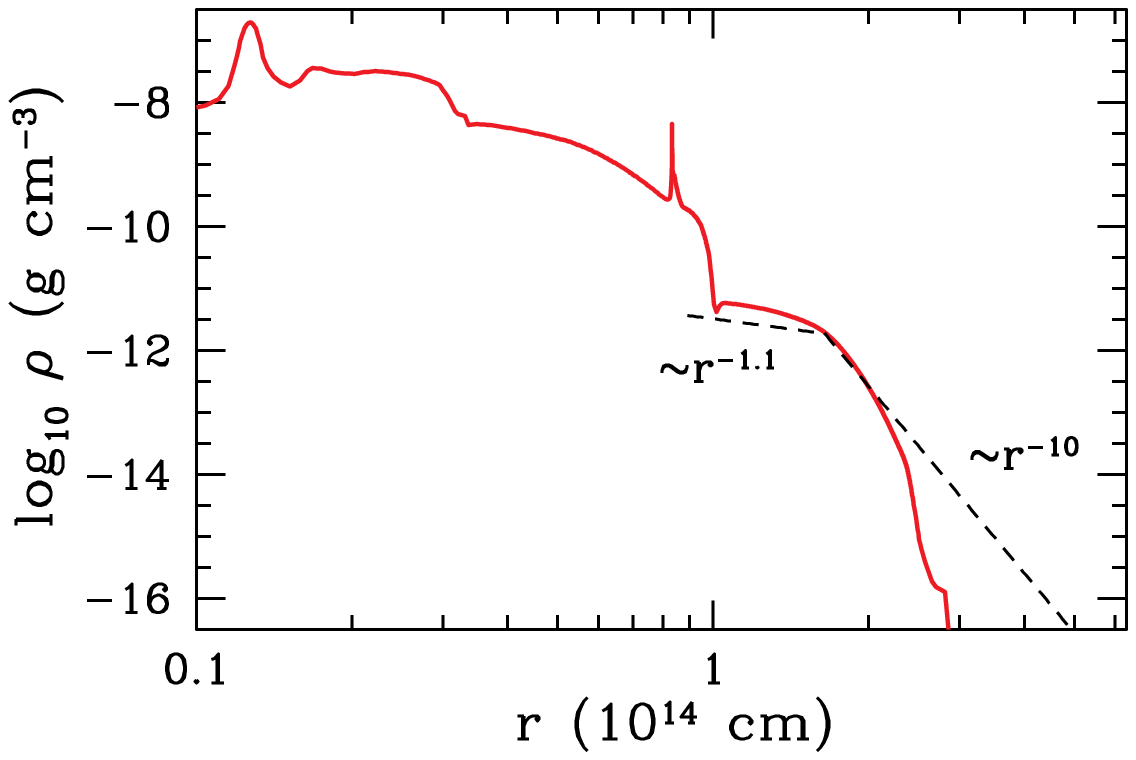}
\includegraphics[width=0.335\textwidth,trim=0.0cm 6cm 0.0cm 0.0cm]{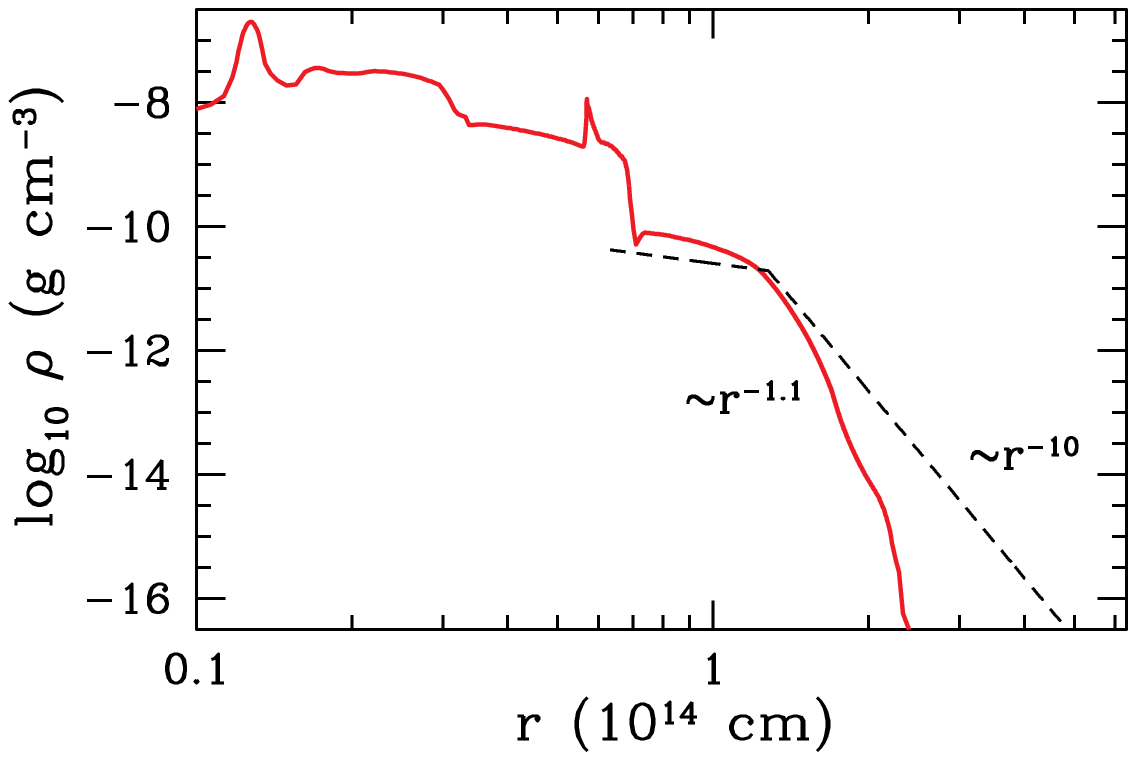}
\caption{Comparison of the density profiles of three additional numerical models taken at $1\,{\rm day}$ following explosion. These were chosen to sample a range of $M_e$ and $R_e$ values, These specifically correspond to $M_e=0.0077\,M_\odot$, $R_e=100\,R_\odot$, $v_t=2.38\times10^9\,{\rm cm\,s^{-1}}$ (left panel), $M_e=0.035\,M_\odot$, $R_e=200\,R_\odot$, $v_t=1.90\times10^9\,{\rm cm\,s^{-1}}$ (center panel), and $M_e=0.17\,M_\odot$, $R_e=125\,R_\odot$, $v_t=1.48\times10^9\,{\rm cm\,s^{-1}}$ (right panel). These values of $v_t$ are derived by combining Equations~(\ref{eq:vt}) and (\ref{eq:e_e}) with $E_{\rm SN}=10^{51}\,{\rm erg}$. In each case, the dashed lines correspond to the analytic density model using Equations~(\ref{eq:rho_out}) and (\ref{eq:rho_in}).}
\label{fig:more rho}
\end{figure}

To test our analytic models, we run numerical SN simulations of helium cores surrounded by hydrogen-rich extended material. A similar model to these ones was the focus of Section~\ref{fig:numerical}, but here we present additional models and discuss the numerical methods in more detail.

These calculations are similar to the work of \citet{Piro17}. We start with a helium core that was generated from a $15\,M_\odot$ zero-age main-sequence star using the 1D stellar evolution code \texttt{MESA} \citep{Paxton13}. Using the overshooting and mixing parameters recommended by \citet{Sukhbold14}, the star is evolved until a large entropy jump between the core and envelope was established. The convective envelope is removed to mimic mass loss during a common envelope phase. The resulting helium core has a mass of $\approx4.95\,M_\odot$. Above the helium core we stitch a low mass hydrogen envelope with a $\rho\propto r^{-3/2}$ density profile to represent the extended material. This specific scaling is meant to mimic the expected profile for a convective layer, but as shown in \citep{Piro17}, SCE is fairly insensitive to this exact choice as long as $M_e$ and $R_e$ are the same.

These models are then exploded with our open-source numerical code \texttt{SNEC} \citep{Morozova15}. We assume that the inner $1.4\,M_\odot$ of the models form a neutron star and excise this region before the explosion. A $^{56}$Ni mass of $0.1\,M_\odot$ is placed at the inner edge of the ejecta, with the exact value not being critically important because we only compare the SCE phase of the simulations with our analytic model.  We use a ``thermal bomb mechanism'' for the explosion, where a luminosity is provided to the inner $0.01\,M_\odot$ of the model for a duration of $0.01\,{\rm s}$ to generate an explosion with energy $10^{51}\,{\rm erg}$.

\begin{figure}
\includegraphics[width=0.34\textwidth,trim=0.3cm 0.3cm 1.3cm 0.0cm]{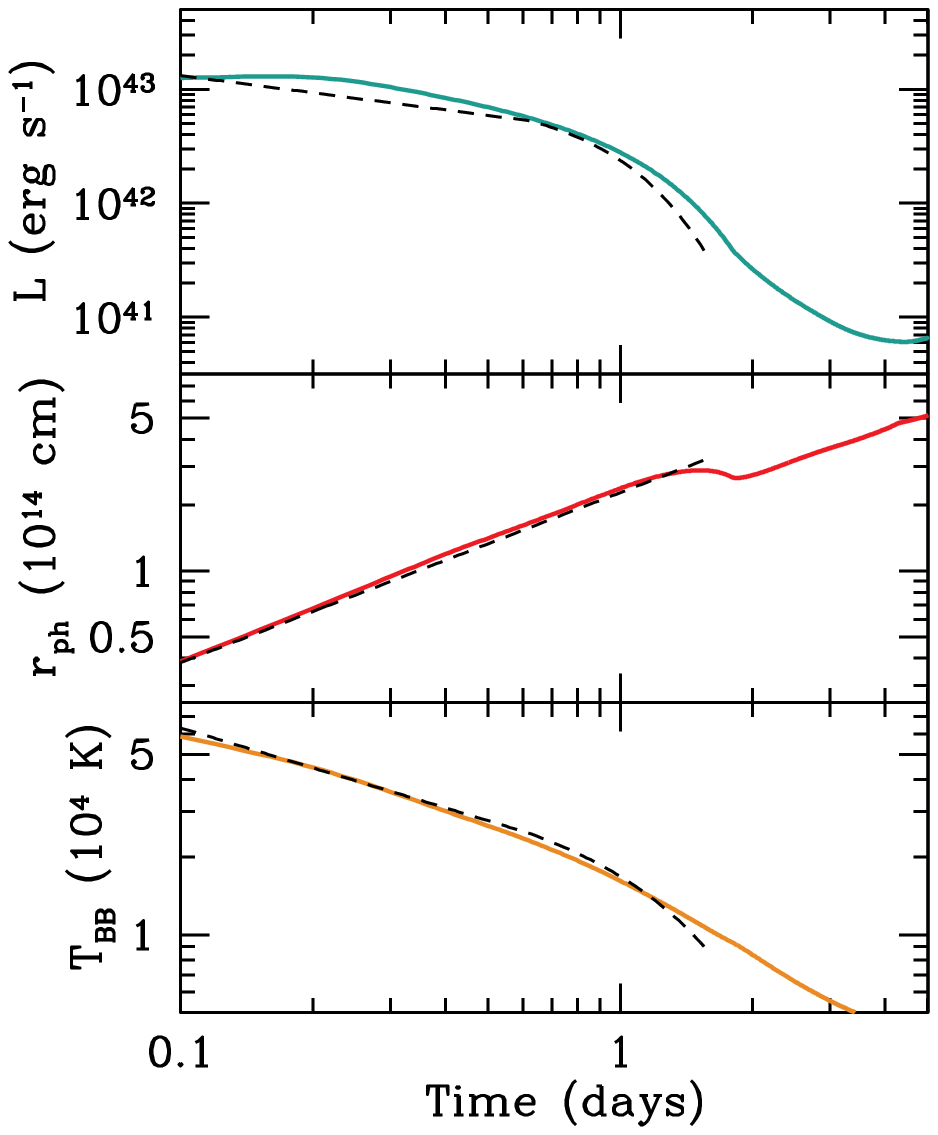}
\includegraphics[width=0.34\textwidth,trim=0.3cm 0.3cm 1.3cm 0.0cm]{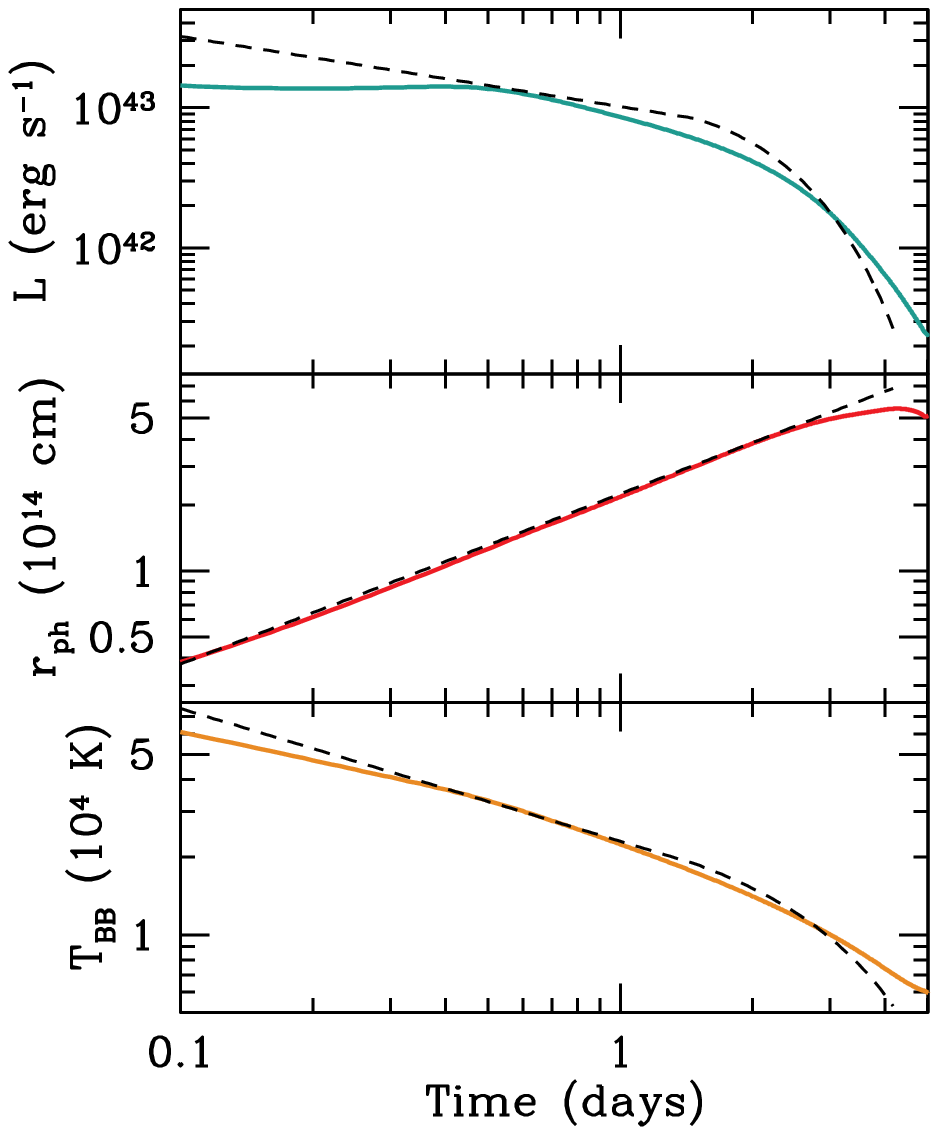}
\includegraphics[width=0.34\textwidth,trim=0.3cm 0.3cm 1.3cm 0.0cm]{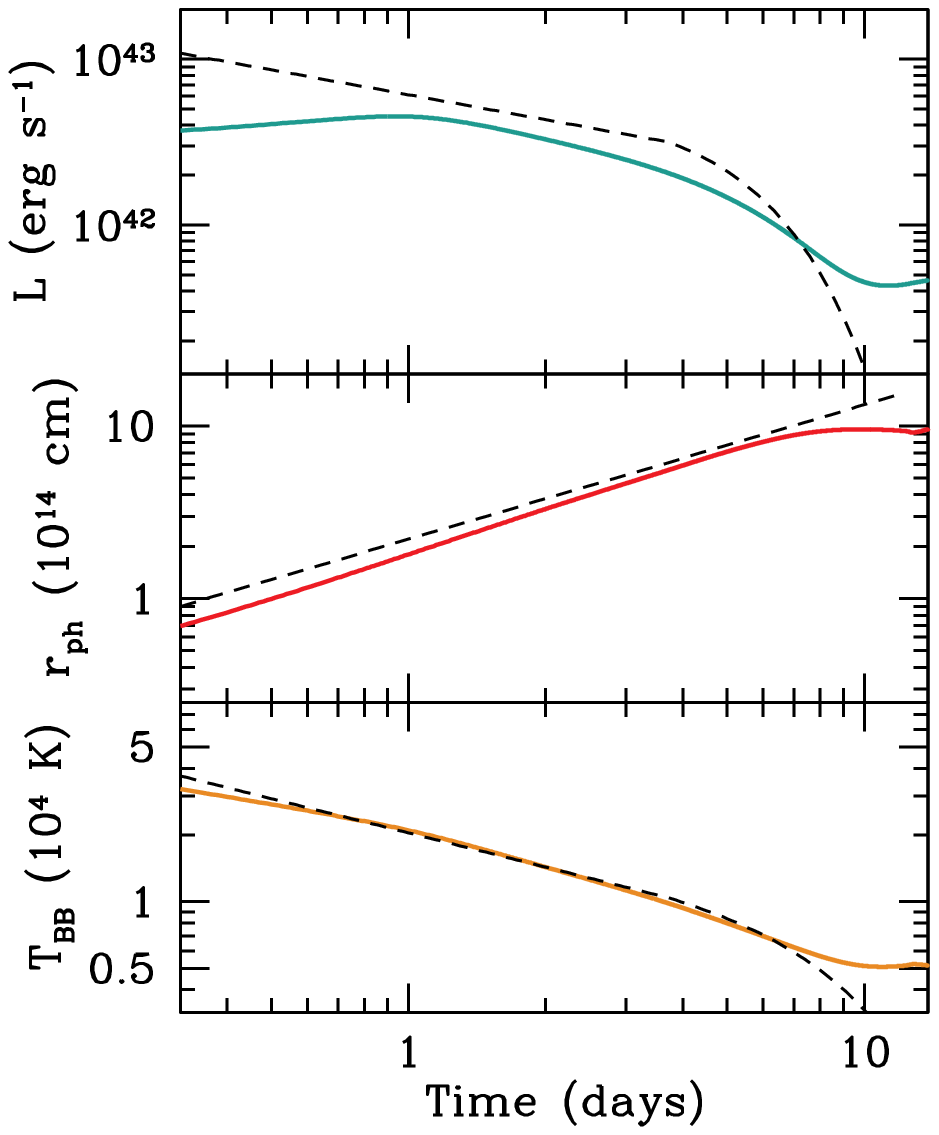}
\caption{Numerical results for the luminosity (turquoise), photospheric radius (red), and temperature (orange) for the same models as Figure~\ref{fig:more rho}. Dashed lines are the analytic models using the same values of $M_e$, $R_e$, and $v_t$.}
\label{fig:more lcs}
\end{figure}

In Figures~\ref{fig:more rho} and \ref{fig:more lcs}, we consider three additional numerical models beyond what is presented in Section~\ref{sec:numerical}. These are chosen to span a range of values for $M_e$, $R_e$, and $v_t$. The dashed lines in Figure~\ref{fig:more rho} show our analytic density model at one day following explosion, again demonstrating that it provides a similar but not exact match to the numerical density profiles. The observables are summarized in Figure~\ref{fig:more lcs}, for which again the analytic models (dashed lines) show reasonable agreement. Perhaps the largest difference is for the right-most panel, which corresponds to the largest extended material mass. This may indicate that a more careful treatment of the relation between $E_e$ and $E_{\rm SN}$, rather than Equation~(\ref{eq:e_e}), may be needed.

\end{appendix}

\bibliographystyle{yahapj}

\end{document}